\documentclass[prb,aps,twocolumn,superscriptaddress,showpacs,nolongbibliography]{revtex4-2}

\usepackage[none]{hyphenat}
\usepackage{amsmath}
\usepackage{amssymb}
\usepackage{color}
\usepackage{feynmp-auto}
\usepackage{graphicx}
\usepackage{grffile}
\graphicspath{{images/}}
\graphicspath{{Figures/}}

\usepackage{tikz}

\begin{document}
\title{Disorder effects in the $\mathbb{Z}_{3}$ Fock parafermion chain}

\author{G.~Camacho}
\affiliation{Technische Universit\"at Braunschweig, Institut f\"ur Mathematische Physik, Mendelssohnstrasse 3, 38106 Braunschweig, Germany}
\author{J. Vahedi}
\affiliation{Department of Physics and Earth Sciences, Jacobs University Bremen, Bremen 28759, Germany}
\author{D. Schuricht}
\affiliation{Institute for Theoretical Physics, Center for Extreme Matter and Emergent Phenomena, Utrecht University, Princetonplein 5, 3584 CC Utrecht, The Netherlands}
\author{C.~Karrasch}
\affiliation{Technische Universit\"at Braunschweig, Institut f\"ur Mathematische Physik, Mendelssohnstrasse 3, 38106 Braunschweig, Germany}

\setlength{\abovedisplayskip}{4pt}
\setlength{\belowdisplayskip}{4pt}  

\date{\today}

\begin{abstract}
We study the effects of disorder in a one-dimensional model of $\mathbb{Z}_{3}$ Fock parafermions which can be viewed as a generalization of the prototypical Kitaev chain. Exact diagonalization is employed to determine level statistics, participation ratios, and the dynamics of domain walls. This allows us to identify ergodic as well as finite-size localized phases. In order to distinguish Anderson from many-body localization, we calculate the time evolution of the entanglement entropy in random initial states using tensor networks. We demonstrate that a purely quadratic parafermion model does not feature Anderson but many-body localization due to the  nontrivial statistics of the particles. 
\end{abstract}

\maketitle

\section{Introduction}\label{sec:intro}

It is well known that low-dimensional quantum systems can host particles with statistical properties beyond the usual boson and fermion paradigms such as anyonic exchange \cite{Wilczek82,Wilczek82prl1} and generalized exclusion statistics \cite{Haldane91prl2,Wu94}. A particular type of particles with anyonic properties is parafermions \cite{FradkinKadanoff80,Fendley12}, which can be viewed as a generalization of the nowadays well-known Majorana fermions \cite{Kitaev_2001}. The latter can be interpreted as real and imaginary parts of spinless fermions, which in turn possess the properties usually attributed to quantum particles such as the existence of a Fock space and therefore a well-defined particle number. The corresponding objects obtained from parafermions were introduced by Cobanera and Ortiz \cite{Cobanera2014,Cobanera17} and named Fock parafermions. By construction, they possess a Fock space with an occupation number, but, in contrast to spinless fermions, also anyonic exchange statistics and a generalized Pauli principle inherited from the underlying parafermion operators.

While Fock parafermions have been utilized to link parafermions to ordinary electrons and thereby investigate models possessing zero-energy (edge) modes \cite{Calzona-18}, they can also be studied as particles in their own right. The first step in this direction was taken by Rossini \emph{et al.}~\cite{Rossini2019}, who studied a tight-binding chain of Fock parafermions. A key observation was that such a model is nonintegrable despite being quadratic in terms of the Fock parafermion operators, and the low-energy properties for generic filling fractions are described by a Luttinger liquid. Due to the generalized Pauli principle, more than one Fock parafermion can occupy a lattice site; coherent pair-hopping processes thus become possible, which yields further gapless phases \cite{Mahyaeh2020}. Very recently, the effect of dissipation was also analyzed \cite{Mastiukova22} which, under suitable conditions, leads to the emergence of a noninteracting single-particle spectrum and dark states. 

Despite these efforts, many open questions on Fock parafermion systems remain, with some of the most natural ones being related to the addition of disorder. It is well known that in other low-dimensional systems, the addition of a disordered potential generally leads to the localization of quantum states and thus drastically affects the transport properties. In noninteracting systems, Anderson localization \cite{Anderson1958,EversMirlin08} manifests itself in a complete freezing of the dynamics, detectable, for example, in the saturation of the entanglement entropy. The generalization of this phenomenon to interacting systems, nowadays known as many-body localization (MBL) \cite{Gornyi2005,Basko2006}, has attracted tremendous attention \cite{Abanin2019}. MBL provides a generic realization of a nonergodic quantum system with potential applications in quantum information \cite{Banuls2017,Friesdorf2015}, and MBL phases feature interesting properties such as an area law \cite{Eisert2010} scaling of the entanglement in the excited states  \cite{Bauer2013}, unconventional transport \cite{Agarwal2017,Luitz2017,BarLev2017}, or a logarithmic growth of the entanglement entropy following quantum quenches \cite{Znidaric2008,Bardarson2012,Serbyn2013}.

In this work, we study disorder effects on Fock parafermions. In particular, we ask whether a purely quadratic Fock parafermion chain with random on-site potential exhibits the phenomenology of Anderson or many-body localization. Our results are consistent with the MBL phenomenology, except in a special limit where the model becomes equivalent to a free fermionic system. Our findings hence suggest that anyonic statistics precludes Anderson localization even in quadratic systems. Furthermore, our results show that the coherent pair hopping supports localization.

Many studies of MBL employ an exact diagonalization of small systems \cite{Pal2010,Luitz2015}. The existence of the MBL phase in the thermodynamic limit has recently been questioned \cite{PhysRevB.100.104204,PhysRevE.102.062144,PhysRevB.103.024203,PhysRevE.104.054105,PhysRevLett.127.230603}, but no conclusive picture has emerged yet \cite{PhysRevLett.124.186601,Panda_2020,PhysRevB.102.100202,ABANIN2021168415,PhysRevB.105.144203,PhysRevB.105.174205}. A possible new viewpoint is quantum avalanches \cite{PhysRevLett.119.150602,PhysRevB.95.155129,PhysRevLett.121.140601,PhysRevB.99.195145,PhysRevB.100.115136,PhysRevResearch.2.033262,PhysRevB.106.L020202,https://doi.org/10.48550/arxiv.2012.15270}. In our work, we address the phenomenology of MBL in finite-size parafermion systems in analogy to the finite-size studies of MBL in the prototypical Heisenberg chain \cite{Pal2010,Luitz2015}.

This article is organized as follows: In the next section, we introduce Fock parafermions, the one-dimensional model we are considering, and recapitulate its basic properties. Thereafter (Sec.~\ref{sec:Methods}), we introduce our numerical approaches, i.e., exact diagonalization (ED) as well as the time-evolving block decimation (TEBD). In Sec.~\ref{sec:Results}, we present our results on the level spacing statistics, the participation ratio, the imbalance dynamics, and the entanglement entropy. In Sec.~\ref{sec:conclusion}, we summarize our findings.

\section{Model}\label{sec:model}
\subsection{Fock parafermions}\label{subsec:theory}

In one dimension, the transverse-field Ising chain is a cornerstone model for the classification of topological order; it is directly linked to the Kitaev chain~\cite{Kitaev_2001,Fendley12} with Majorana edge zero modes appearing in its ordered phase. More generally, one can study $p$-state clock models, and the simplest extension of the Ising chain ($p=2$) is given by the quantum Potts model,
\begin{eqnarray}\label{eq:Potts_model}
H_{\text{Potts}}=&-&J\sum_{j}\left(\sigma_{j}^{\dagger}\sigma_{j+1} + \sigma_{j+1}^{\dagger}\sigma_{j}\right)\nonumber\\ &-&f\sum_{j}\left(\tau_{j}^{\dagger}+\tau_{j}\right),
\end{eqnarray}
where the clock matrices $\sigma_{j},\tau_{j}$ satisfy the following algebra:
\begin{eqnarray}\label{eq:sigma_matrices}
    \sigma_{j}^{\dagger}&=&\sigma_{j}^{p-1},\hspace{5pt} \tau_{j}^{\dagger} =\tau_{j}^{p-1},\nonumber\\
    \sigma_{j}^{p} &=&\tau_{j}^{p}=1,\hspace{5pt}\sigma_{j}\tau_{k}=\omega^{\delta_{j,k}}\tau_{k}\sigma_{j},
\end{eqnarray}
with $\omega=\exp(2\pi\mathrm{i}/p)$. Their explicit representation reads 
\begin{eqnarray}\label{eq:sigma_tau_matrix_elements}
\left[\sigma_j\right]_{kl} &=& \delta_{k+1,l} + \delta_{k,p}\delta_{l,1},\nonumber\\
\left[\tau_j\right]_{kl} &=&\omega^{k-1}\delta_{k,l},\hspace{15pt} k,l\in\{1,...,p\}.
\end{eqnarray}
The notion of a Majorana is then generalized by introducing two parafermion operators at each site $j$, $\gamma_{2j-1}$ and $\gamma_{2j}$, via
\begin{eqnarray}\label{eq:sigma_gamma_mapping}
    \gamma_{2j-1}=\left(\prod_{k<j}\tau_{k}\right)\sigma_{j},\hspace{5pt}\gamma_{2j}=\omega^{\frac{p-1}{2}}\gamma_{2j-1}\tau_{j},
\end{eqnarray}
which satisfy the following algebraic relations
\begin{eqnarray}\label{eq:gamma_comms}
\gamma_{j}\gamma_{k}=\omega^{\text{sgn}(k-j)}\gamma_{k}\gamma_{j},~~
\gamma_{j}^{p-1}=\gamma_{j}^{\dagger},\hspace{10pt}\gamma_{j}^{p}=1.
\end{eqnarray} 
For $p=2$, Eq.~\eqref{eq:gamma_comms} reduces to the usual anticommutation relations for $2L$ Majorana fermions.

The biggest drawback of using the parafermion operators defined in Eq.~\eqref{eq:sigma_gamma_mapping} is that $\gamma_{2j-1}^{\dagger}$ and $\gamma_{2j}^{\dagger}$ cannot be interpreted as particle creation operators. However, it was shown that for any set of parafermion operators governed by Eq.~(\ref{eq:gamma_comms}), a generalized set of annihilation operators $F_j$, the so-called Fock parafermions, can be defined via \cite{Cobanera2014}
\begin{eqnarray}
F_{j}=\frac{p-1}{p}\gamma_{2j-1} -\frac{1}{p}\sum_{m=1}^{p-1}\omega^{\frac{m(m+p)}{2}}\gamma_{2j-1}^{m+1}\gamma_{2j}^{\dagger m}.
\end{eqnarray} 
These operators feature anyonic commutators
\begin{eqnarray}\label{eq:FjFk_rels}
F_{j}F_{k} &=&\omega^{\text{sgn}(k-j)}F_{k}F_{j}, \nonumber\\
F_{j}^{\dagger}F_{k} &=&\omega^{-\text{sgn}(k-j)}F_{k}F_{j}^{\dagger},
\end{eqnarray}
and satisfy the local relations
\begin{eqnarray}\label{eq:F_j_rels}
F_{j}^{p}=0,\hspace{10pt}F_{j}^{\dagger m}F_{j}^{m} + F_{j}^{p-m}F_{j}^{\dagger(p-m)}=1,
\end{eqnarray}
with $m=1,\ldots,p-1$. Note that for $p=2$, this scheme reduces to the standard representation of Majoranas in terms of spinless fermions. The Fock parafermion operators act on the occupation basis of a Fock space in the usual way, thus they can be interpreted as annihilating and creating particles that satisfy anyonic statistics. In particular, the occupation number basis of the Fock space is obtained by repeated application of the creation operators over the vacuum $|0\rangle$,
\begin{eqnarray}\label{eq:nocc_stateF}
|n_{1},n_{2},...,n_{L}\rangle=F_{1}^{\dagger n_{1}}F_{2}^{\dagger n_{2}}...F_{L}^{\dagger n_{L}}|0\rangle,
\end{eqnarray} 
with $n_k\in\{0,1,\ldots,p-1\}$, and $L$ denoting the total number of lattice sites. Due to Eq.~\eqref{eq:F_j_rels}, each lattice site can accommodate, at most, $p-1$ Fock parafermions. One can easily show that the states (\ref{eq:nocc_stateF}) indeed form an orthonormal basis,
\begin{eqnarray}\label{eq:nocc_state}
\langle n_1,\ldots,n_L |m_{1},...,m_{L}\rangle =\delta_{n_1,m_1}\cdots\delta_{n_L,m_L}.
\end{eqnarray} 
The number operator at a given lattice site $j$ is given by 
\begin{eqnarray}\label{eq:Nj_ops}
N_{j}=\sum_{m=1}^{p-1}F_{j}^{\dagger m}F_{j}^{m},
\end{eqnarray}
which satisfies the commutation relations
\begin{eqnarray}\label{eq:commutations}
\big[N_{j},F_{j}^{\dagger}\big] =F_{j}^{\dagger},~~~
\big[N_{j},F_{j}\big] =-F_{j}.
\end{eqnarray}
This entails, in particular,
\begin{eqnarray}
N_j|n_{1},n_{2},...,n_{L}\rangle=n_j|n_{1},n_{2},...,n_{L}\rangle .
\end{eqnarray} 

In analogy with conventional fermions, the application of $F_j^\dagger$ to a basis state $|n_{1},...,n_j,...,n_{L}\rangle$ yields a statistical phase. In order to facilitate the implementation of exact diagonalization and tensor network techniques, we employ the Fradkin--Kadanoff (generalized Jordan--Wigner) transformation~\cite{FradkinKadanoff80},
\begin{eqnarray}\label{eq:FK_transf}
F_{j}=\left(\prod_{l=1}^{j-1}U_{l}\right)B_{j}.
\end{eqnarray}
In the Fock basis, the operators $U_{j},B_{j}$ are explicitly given by
\begin{eqnarray}\label{eq:UB_matrices}
B_j|n_{1},...,n_j,...,n_{L}\rangle &=& |n_{1},...,n_j-1,...,n_{L}\rangle, \nonumber \\
U_j|n_{1},...,n_j,...,n_{L}\rangle &=& \omega^{n_j}|n_{1},...,n_j,...,n_{L}\rangle,
\end{eqnarray} 
which entails that they commute on different sites.

\subsection{Hamiltonian}\label{subsec:hamiltonian}
From now on, we restrict ourselves to $p=3$, i.e., the simplest case showing nontrivial anyonic statistics. The Potts model does not take a simple form when expressed in terms of Fock parafermion operators. Instead, we start with a model of Fock parafermions directly. Following prior works \cite{Rossini2019,Mahyaeh2020}, we study the following Hamiltonian:
\begin{eqnarray}\label{eq:H}
H_{0} &=&-J\sum_{j=1}^{L-1}\left[(1-g)F_{j}^{\dagger}F_{j+1} + gF_{j}^{\dagger 2}F_{j+1}^{2}+ \text{H.c.}\right],\nonumber\\
H &=& H_0 + \sum_{j=1}^{L}\mu_{j}N_{j},
\end{eqnarray}
where $g\in[0,1]$ tunes between pure single-particle and coherent pair hopping. We will always use $J=1$ as the unit of energy. The disordered on-site potentials $\mu_j$ are taken from a uniformly sampled distribution, $\mu_{j}\in\left[-W,W\right]$, with $W$ representing the disorder strength.  We consider open boundary conditions. After the Fradkin--Kadanoff transformation, the Hamiltonian takes the form \cite{Rossini2019,Mahyaeh2020}
\begin{eqnarray}\label{eq:mapped_H_UB}
    H_0 &=&-J\sum_{j=1}^{L-1}\left[(1-g)B_{j}^{\dagger}U_{j}B_{j+1}+gB_{j}^{\dagger 2}B_{j+1}^{2}+\text{H.c.}\right],\nonumber\\
    H &=& H_0 + \sum_{j=1}^{L}\mu_{j}\left(B_{j}^{\dagger}B_{j}+B_{j}^{\dagger 2}B_{j}^{2}\right).
\end{eqnarray}

The model at $g=W=0$ corresponds to a tight-binding chain of Fock parafermions \cite{Rossini2019}. We note that even though the model is quadratic, the system is nevertheless nonintegrable due to the nontrivial statistics of the particles, which prohibits the application of Wick's theorem \cite{Rossini2019}. At $g\neq 0$, a coherent pair hopping is included, the influence of which  was studied in Ref.~\onlinecite{Mahyaeh2020} (for $W=0$).

The Hamiltonian in Eq.~\eqref{eq:H} features two key symmetries, namely, (i) a $U(1)$ symmetry related to the conservation of the total particle number $N=\sum_{j}N_{j}$, and (ii) a particle-hole symmetry in the case $p=3$ \cite{Mahyaeh2020}. Thus, we can work in sectors with fixed $N$ and, moreover, restrict ourselves to filling fractions $n=N/L\in\left[0,1\right]$. In our work, we will always consider $n=0.5$.

The phase diagram at $W=0$ and arbitrary values of the filling $n$ and coupling $g$ was studied in Ref.~\onlinecite{Mahyaeh2020} by numerical and analytical approaches. At $n=0.5$, the system exhibits two gapless phases separated by a (second-order) transition around $g_\mathrm{c}\sim 0.6$. Both phases display Luttinger liquid characteristics, i.e., they can be described by a conformal field theory with a central charge $c=1$. The behavior of correlation functions such as $\langle F_{i}^\dagger F_j\rangle$, however, differs qualitatively in the two phases. 

A simplification occurs in the case $g=1$. As can be seen from Eq.~(\ref{eq:H}), sites with one Fock parafermion per lattice site completely decouple, i.e., singly occupied sites cease to move. The full chain is thus broken into segments of empty and doubly occupied sites which are separated by singly occupied sites. On these segments, the local on-site basis states of $|0\rangle,|2\rangle$ can be identified with a spin-1/2 degree of freedom, and the Hamiltonian becomes equivalent to an \emph{XX} spin chain, which is well known to be noninteracting \cite{Lieb-61}. Hence, at $g=1$, the spectrum of the full chain becomes equivalent to a collection of free systems.

\begin{figure}[t!]
\includegraphics[scale=0.5]{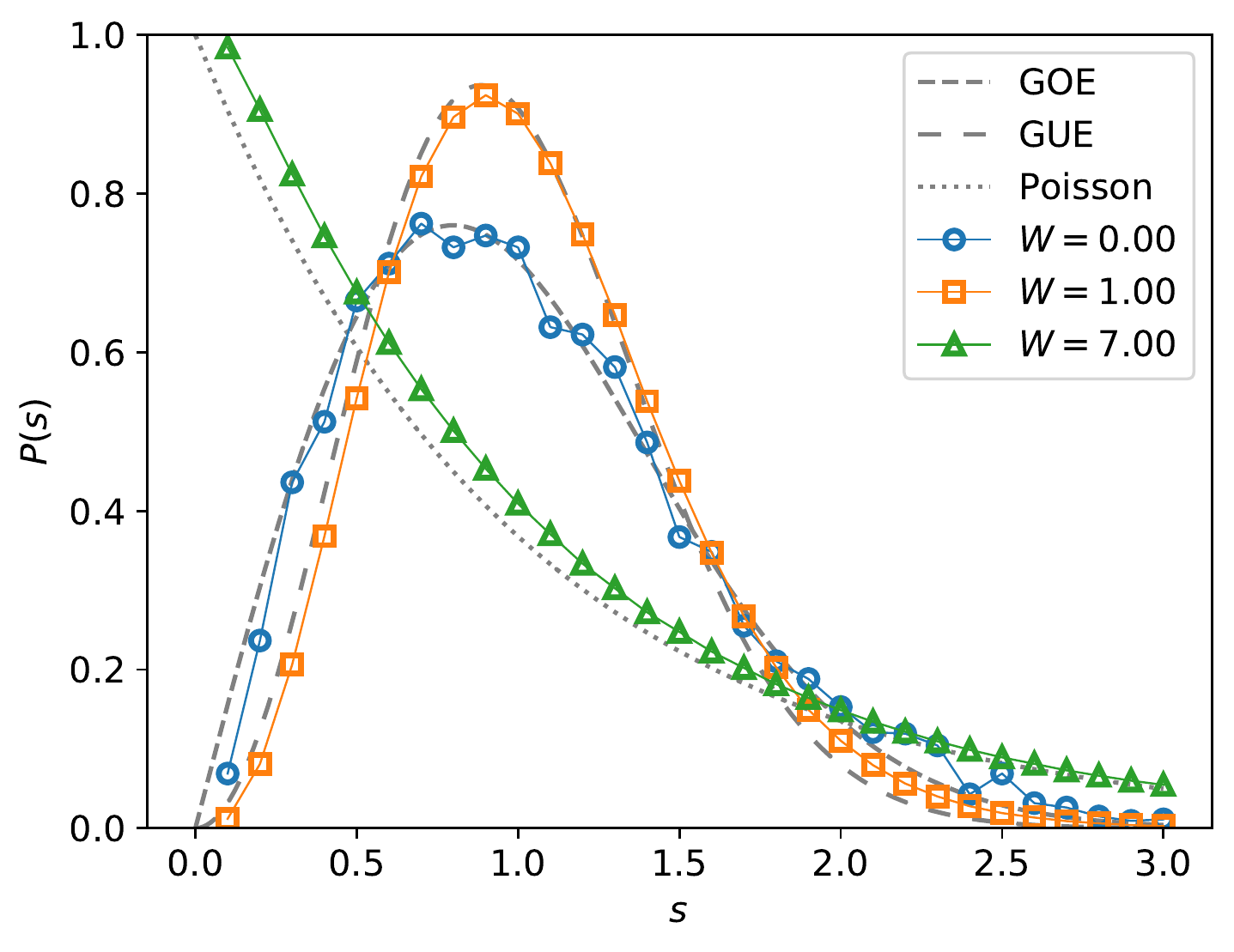}
\caption{Distribution $P(s)$ of the levels statistics in different disorder regimes at $g=0.5$. The data were obtained by a full ED calculation of the spectrum at $L=12$. For $W=0.0$, the distribution follows the GOE. A small $W=1.0$ breaks the discrete $\mathbb{Z}_2$ symmetry, and the level spacing statistics follows the GUE. For high values of $W=7.0$, the level spacing follows Poissonian statistics. For finite disorder $W>0$, $P(s)$ was obtained by averaging over 1000 samples.} 
\label{fig:figure_Ps}
\end{figure}

\section{Methods}\label{sec:Methods}

\subsection{Exact diagonalization}\label{subsec:ED}
Due to the exponential increase of the Hilbert space dimension with the system size, ED methods are limited to the study of small systems. In this work, we use full ED to study chains of up to $L=14$ sites with open boundary conditions.

Calculating all eigenstates in a sector with a given filling fraction $n$ often becomes computationally expensive, in particular if one wants to average over a substantial number of disorder realization. It is then beneficial to target a specific energy density $\epsilon$ within the spectrum and to restrict the calculation of eigenstates to a small number around this value. To this end, we employ the shift-invert method, which has been applied previously in the context of Anderson localization~\cite{Rodriguez2011} and, more recently, in disordered quantum spin chains~\cite{Luitz2015}. At any disorder realization, we calculate the minimum (maximum) eigenvalue of the spectrum $E_{0}$ ($E_{\text{max}}$), which allows us to define the normalized energy density $\epsilon=(E-E_{\text{max}})/(E_{0}-E_{\text{max}})$~\cite{Luitz2015}. A set of 20 to 30 pairs of eigenenergies is then calculated around a specific value of $\epsilon$.

The shift-invert technique allows us to compute observables in an energy-resolved way at a much reduced numerical effort. All ED results, except those in Fig.~\ref{fig:figure_Ps}, were obtained using this approach. We checked that for small system sizes, using a full ED yields similar results.

In Sec.~\ref{subsec:imbalance}, we will study the time evolution of initial product states. This is accomplished via Krylov space techniques~\cite{Paeckel2019}.

\subsection{Tensor networks}\label{subsec:DMRG}
As a second approach, we employ tensor network methods to simulate the time evolution of given initial states~\cite{Schollwoeck2011}. In particular, we use the TEBD with a fourth-order Suzuki-Trotter expansion and a time step of $dt=0.1$. The bond dimension is dynamically increased in order to maintain a fixed discarded weight, which we varied over two orders of magnitude ($10^{-9}-10^{-7}$) in order to check for convergence. (We have also checked our TEBD approach against data obtained via the Krylov time-evolution method.)

\begin{figure}[t!]
\includegraphics[scale=0.28]{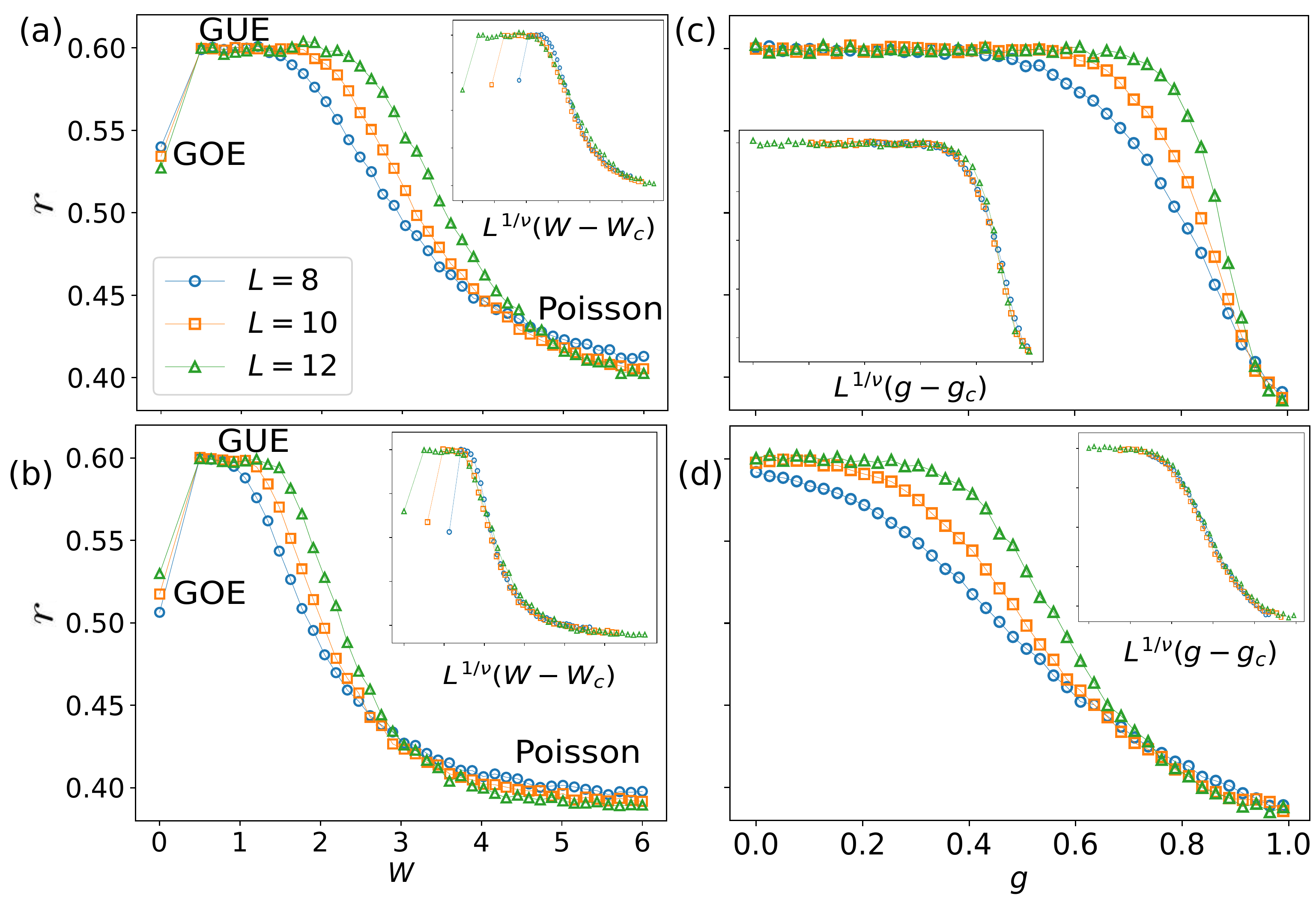}
\caption{Adjacent gap ratio for different system sizes and mid-spectrum states at (a) fixed $g=0.2$, (b) fixed $g=0.5$, (c) fixed $W=1$, and (d) fixed $W=2$. The analytically known values~\cite{Atas-13} in the different regimes are given by $r_{\text{GOE}}\sim 0.5307$, $r_{\text{GUE}}\sim 0.5996$, and $r_{\text{Poisson}}\sim 0.3863$. \textit{Insets}: In order to quantify the crossover between ergodic and localized phases, the data are collapsed by rescaling the $x$ axes as $L^{1/\nu}\left(W-W_{c}\right)$ or $L^{1/\nu}\left(g-g_{c}\right)$, which yields $\nu=0.75$ and (a) $W_{c}=4.2$, (b) $W_{c}=2.7$, (c) $g_{c}=0.97$, and (d) $g_{c}=0.7$.} 
\label{fig:figure_2}
\end{figure}

\section{Results}\label{sec:Results}
\begin{figure}[t]
\includegraphics[scale=0.3]{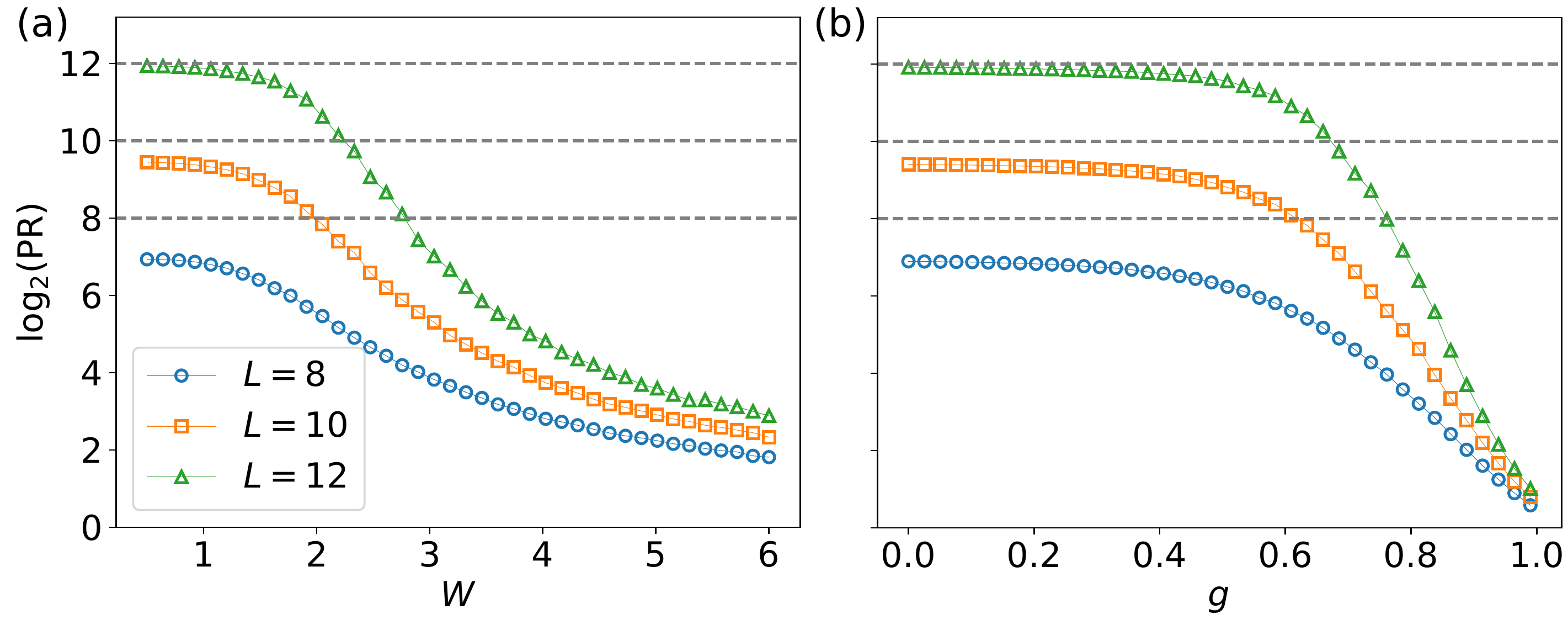}
\caption{Participation ratio for different system sizes and mid-spectrum states at (a) fixed $g=0.2$, and (b) fixed $W=1$. The ergodic and localized regimes are governed by $\text{PR}\sim 3^L$ and $\text{PR}\sim 1$, respectively. Dashed lines indicate $\sim 3^L$ scaling.}
\label{fig:figure_3}
\end{figure}

\begin{figure}[t]
\includegraphics[scale=0.28]{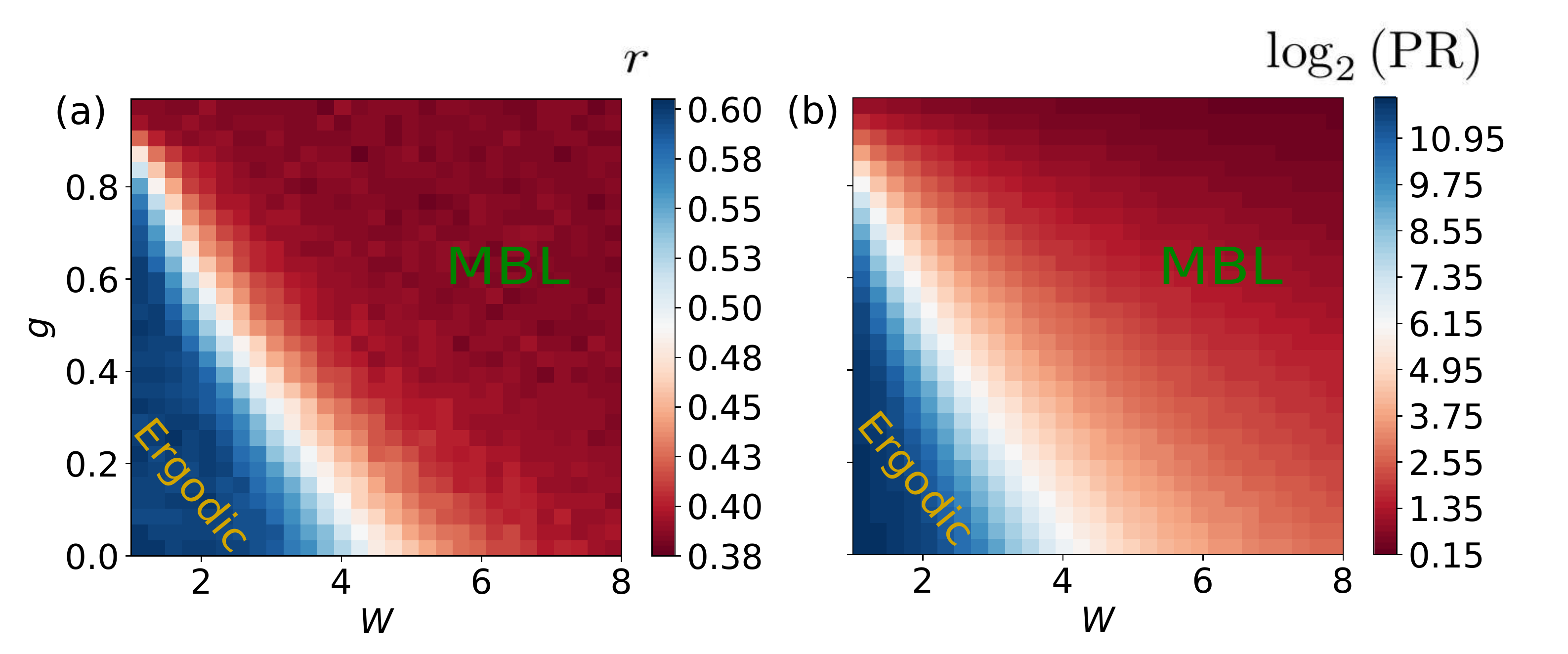}
\caption{(a) Adjacent gap ratio and (b) participation ratio of mid-spectrum states for a system of $L=12$ sites as a function of both the disorder $W$ and the interaction $g$. Blue and red coloring marks ergodic and localized regimes, respectively.}
\label{fig:figure_1}
\end{figure}

We now discuss the effects of disorder in our model. We focus on a set of standard observables. We reiterate that all calculations are carried out for fillings $n=0.5$ and that all ED data, except those in Fig.~\ref{fig:figure_Ps}, were obtained using shift invert. The data at $W>0$ were averaged over $\mathcal{O}(1000)$ disorder realizations \footnote{We used the following sample sizes. Figs.~\ref{fig:figure_Ps}, \ref{fig:figure_1}, \ref{fig:figure_4}, and \ref{fig:figure_6}: 1000 samples. Figs.~\ref{fig:figure_2} and \ref{fig:figure_3}: 3000, 2000, and 1000 samples at $L=8$, $L=10$, and $L=12$, respectively. Fig.~\ref{fig:figure_5}: 500 samples. }.

\subsection{Level spacing statistics}\label{subsec:LSS}
Random matrix theory predicts that the distribution $P(s)$ of adjacent gaps, $s_{n}=E_{n+1}-E_{n}$, in the spectrum of random matrices follows one of the so-called Gaussian ensembles~\cite{Beenakker97}. Purely real Hamiltonians will generally obey the statistics from the Gaussian orthogonal ensemble (GOE), while general hermitian Hamiltonians are governed by the Gaussian unitary ensemble (GUE). For Hamiltonians possessing an infinite set of conservation laws such as integrable or free systems, the distribution $P(s)$ follows a Poisson distribution. The same holds true for MBL systems.

In Fig.~\ref{fig:figure_Ps}, we show results of a full ED calculation for $P(s)$ for different values of the disorder strength $W$. Following Rossini \emph{et al.}~\cite{Rossini2019}, we have discarded the upper and lower third of the spectrum to generate the distributions, and we have not employed any unfolding procedures \footnote{To be precise: We calculate the normalized difference of  eigenvalues $s_n=(E_{n+1}-E_n)/[\sum_n(E_{n+1}-E_n)/N]$, where $n$ runs only over the middle of the spectrum (upper and lower third disregarded) and $N$ is the total number of eigenvalues in this region. We sort $s_n$ into 30 bins of width $0.1$ and plot the number of eigenvalues per bin divided by $0.1N$ (thus disregarding values of $s_n>3$). The analytical functions plotted in Fig.~\ref{fig:figure_Ps} are given by~\cite{Beenakker97} $f_\text{GOE}(s)=\frac{\pi s}{2}\exp(-\pi s^{2}/4)$, $f_\text{GUE}(s)=\frac{32 s^{2}}{\pi^{2}}\exp(-4s^{2}/\pi)$, and $f_\text{Poisson}(s)=\exp(-s)$.}. 

Without disorder, $P(s)$ follows GOE statistics. This can be understood by noting that in the absence of the disordered on-site potential, the model possesses a discrete $\mathbb{Z}_2$ particle-hole symmetry $F_j\to F_j^\dagger$ under which the particle number operator transforms as $N_j\to 2-N_j$ \cite{Mahyaeh2020}. This symmetry is absent for $W>0$, and $P(s)$ is governed by the GUE. For large $W=7$, one obtains a Poissonian distribution. This hints at a crossover between an ergodic and a localized phase, which we will further investigate below. (Note that the case $g=1$ features a Poissonian distribution at any $W$ due to the mapping to a collection of \emph{XX} chains.)

As an alternative to $P(s)$, one can compute the adjacent gap ratio~\cite{Oganesyan2007},
\begin{eqnarray}\label{eq:rn_def}
r= \frac{\min\left(s_{n+1},s_{n}\right)}{\max\left(s_{n+1},s_{n}\right)},
\end{eqnarray}
which, after disorder averaging, takes the value $r_{\text{GOE}}\sim 0.5307$ in the GOE, $r_{\text{GUE}}\sim 0.5996$ in the GUE, and $r_{\text{Poisson}}\sim 0.3863$ in the case of Poissonian statistics \cite{Atas-13}.

In Figs.~\ref{fig:figure_2}(a) and~\ref{fig:figure_2}(b), we show $r$ at $g=0.2$ and $g=0.5$ as a function of $W$ for mid-spectrum states ($\epsilon=0.5$) and different $L$. We can again identify GOE statistics at $W=0$, GUE statistics at small $W$, and Poissonian behavior at large $W$. In order to quantify where the crossover between the ergodic and the localized regime takes place, we have collapsed the data by rescaling the $x$ axis as $L^{1/\nu}\left(W-W_{c}\right)$ (inset). This yields $W_{c}=4.2$ as well as $W_{c}=2.7$ for the critical disorder strength. 

In Figs.~\ref{fig:figure_2}(c) and~\ref{fig:figure_2}(d), we repeat this analysis for fixed $W=1$ and $W=2$. We observe a crossover between GUE and Poissonian behavior as $g$ is increased and obtain $g_{c}=0.97$ and $g_{c}=0.7$ as the critical interaction for the crossover between ergodic and localized regimes. 

One should note that even though the Hamiltonian is purely quadratic in terms of the Fock parafermion operators at $g=0$, the system is not free due to the nontrivial statistics of the particles. In contrast, at $g=1$, the model can be mapped to a collection of noninteracting \emph{XX} spin chains. It is thus plausible that the size of the ergodic regime shrinks with increasing $g$.

In sum, the phenomenology of the level statistics of our disordered parafermion model is analogous to the disordered \emph{XXZ} chain, except for the fact that the ergodic regime is generically governed by a GUE ensemble (instead of a GOE ensemble in the \emph{XXZ} case).

\subsection{Participation ratio}\label{subsec:IPR}

We introduce the inverse participation ratio (IPR) of a state $|\psi_{k}\rangle$ over a given spatial basis $|n\rangle$ as \cite{Misguich2016}
\begin{eqnarray}\label{eq:IPR_k}
\text{IPR}_{k}=\sum_{n=1}^{\mathcal{D}} |\langle n|\psi_{k}\rangle|^{4},
\end{eqnarray} 
with $\mathcal{D}$ representing the Hilbert space dimension. We choose $|\psi_{k}\rangle$ as an eigenstate of $H$ and take $|n\rangle$ as the occupation number basis of Eq.~(\ref{eq:nocc_state}). We determine $m\sim30$ eigenpairs in an energy window around $\epsilon$ using shift invert and define the normalized inverse participation ratio as
\begin{eqnarray}\label{eq:PR}
\text{IPR}(\epsilon)=\frac{\sum_{k=1}^{m}\text{IPR}_{k}}{m}=\frac{\sum_{k=1}^{m}\sum_{n=1}^{\mathcal{D}} |\langle n|\psi_{k}\rangle|^{4}}{m}.
\end{eqnarray}
We first compute the disorder-averaged IPR, and from that, the participation ratio,
\begin{eqnarray}\label{eq:IPR}
\text{PR}(\epsilon)=\frac{1}{\text{IPR}(\epsilon)}.
\end{eqnarray}

In the ergodic regime, any eigenstate takes the form $|\psi_{k}\rangle=\sum_{n}c_{n,k}|n\rangle$ with approximately equal weights, $|c_{n,k}|^{2}=|\langle n|\psi_{k}\rangle|^{2}\sim 1/\mathcal{D}$. This entails $\text{IPR}_{k}\sim 1/\mathcal{D}$ as well as $\text{PR}(\epsilon)\sim \mathcal{D}\sim 3^L$. On the other hand, in a localized phase, eigenstates are close to spatial product states, which leads to $c_{n,k}\sim \delta_{n,n_0}$ and thus $\text{IPR}_{k}\sim 1$ as well as $\text{PR}(\epsilon)\sim1$.

In Figs.~\ref{fig:figure_3}(a) and~\ref{fig:figure_3}(b), we show the participation ratio for mid-spectrum states ($\epsilon=0.5$) and different $L$ at fixed $g=0.2$ and fixed $W=1$, respectively. One can identify the ergodic and localized regimes in qualitative agreement with Fig.~\ref{fig:figure_2} \footnote{Note that even though the total Hilbert space dimension is $\mathcal{D}=3^{L}$, the dimension of the subspace with filling $n=0.5$ is considerably smaller. We thus plot $\text{PR}$ on a log scale with base 2 but emphasize that this has no bearing on our findings.}.

\begin{figure}[t]
\includegraphics[width=0.92\columnwidth]{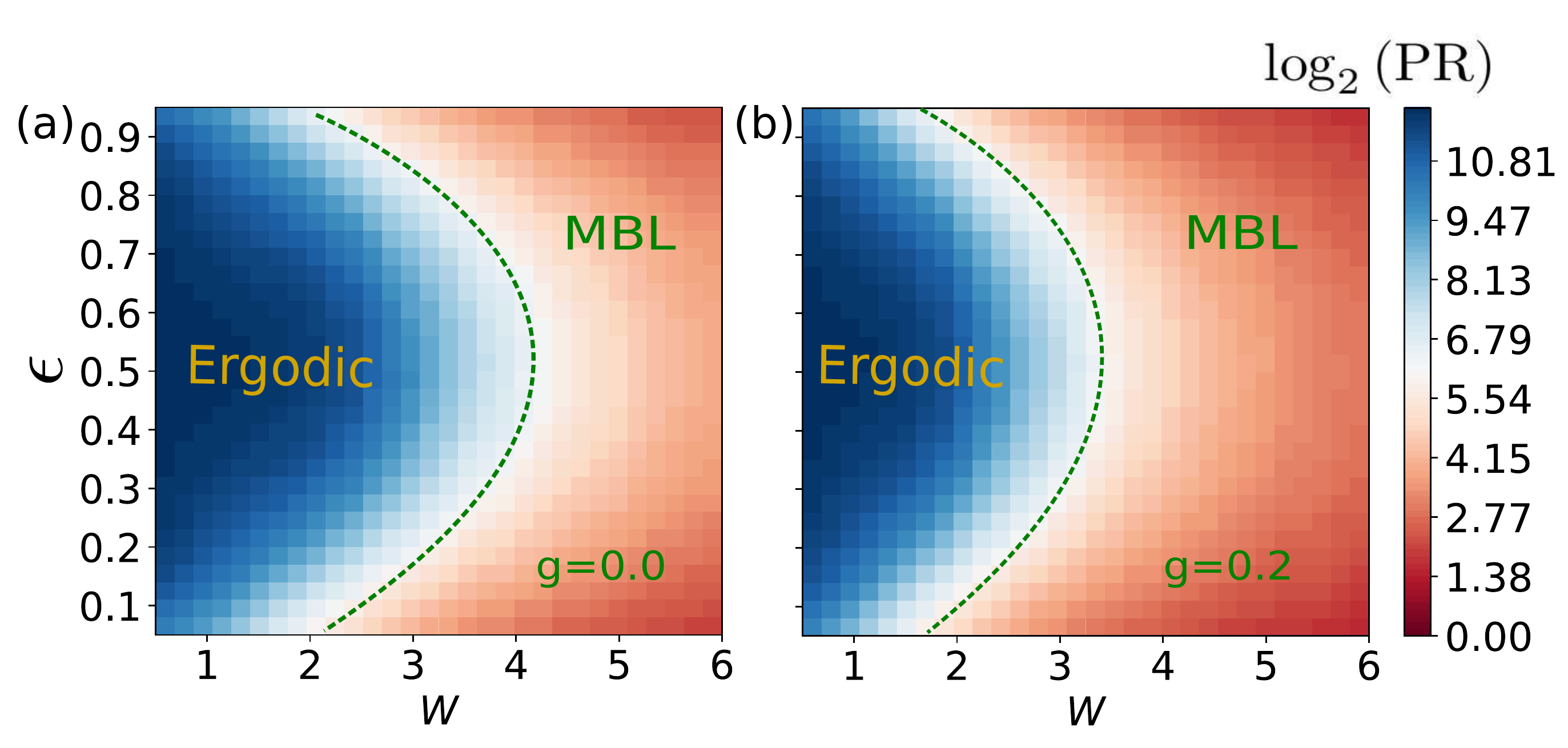}
\caption{Participation ratio as a function of the disorder strength $W$ and the energy density $\epsilon$ for (a) $g=0.0$ and (b) $g=0.2$ on a chain of $L=12$ sites. The green dashed line corresponds to $\log_{2}(\text{PR})=L/2=6$.} 
\label{fig:figure_4}
\end{figure}

\subsection{Phase diagram}

In Fig.~\ref{fig:figure_1}, we show the adjacent gap ratio as well as the participation ratio as a function of both the disorder strength $W$ and the interaction $g$ for mid-spectrum states ($\epsilon=0.5$) of a system of $L=12$ sites. Blue and red coloring indicate ergodic and localized regimes, respectively. One can again see that the size of the localized phase grows with $g$.

In Fig.~\ref{fig:figure_4}, we plot the participation ratio as a function of the disorder and the energy density $\epsilon$ for two values of $g$ at $L=12$. Moving away from mid-spectrum states favors localization in analogy with, e.g., the behavior of the disordered \emph{XXZ} chain \cite{Luitz2015}.

\subsection{Imbalance dynamics}\label{subsec:imbalance}

We complement our analysis of the crossover between ergodic and localized regimes by studying the imbalance dynamics as another prototypical setup. We initially prepare the system in a domain wall state of the form
\begin{equation}\label{eq:wall_state}
    |\psi_{0}\rangle=|1,1,\ldots,1,n_{L/2}=1,0,\ldots,0\rangle.
\end{equation}
The time evolution is calculated using Krylov space methods. We focus on the number of particles transported across the domain,
\begin{eqnarray}\label{eq:imb_def}
\mathcal{I}(t)=2\frac{N_{L}(t)-N_{R}(t)}{L},
\end{eqnarray}
where $N_{L}(t)=\sum_{j=1}^{L/2}\langle\psi(t)|N_{j}|\psi(t)\rangle$, and $N_{R}(t)$ defined analogously.
 
Ergodic systems thermalize; all knowledge from the initial state is eventually lost during the time evolution, which entails $\mathcal{I}(t\to\infty)\to 0$. In contrast, a localized phase features $\mathcal{I}(t\to\infty)>0$. This expectation was confirmed, e.g., for the disordered \emph{XXZ} chain \cite{Luitz2016,PhysRevB.94.161109}. 

\begin{figure}[t!]
\begin{center}
\includegraphics[scale=0.28]{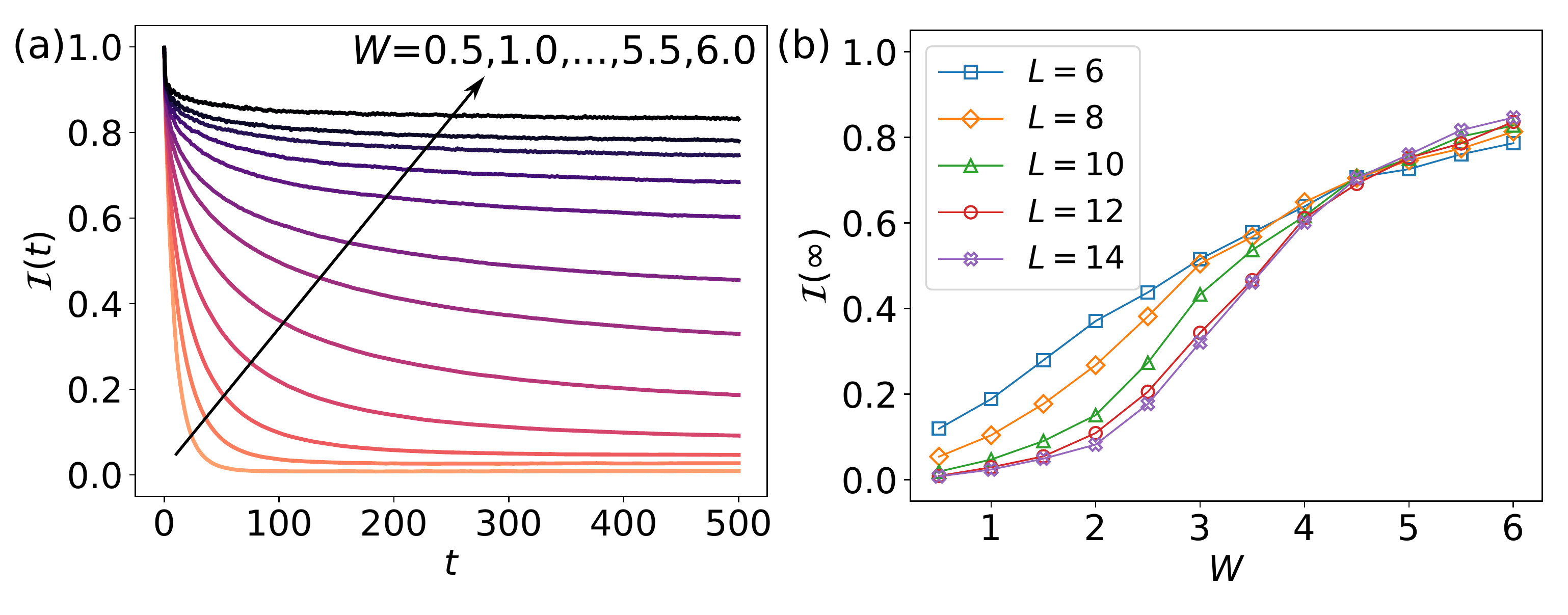}
\caption{(a) Imbalance dynamics for $g=0.2$ and different $W$ for a system of $L=12$ sites. (b) The corresponding steady-state value $\mathcal{I}(\infty)$ as a function of $W$ for various $L$.}
\label{fig:figure_5} 
\end{center}
\end{figure}

In Fig.~\ref{fig:figure_5}(a), we show $\mathcal{I}(t)$ at $g=0.2$ for different $W$ at $L=12$. The curves decay slower for larger $W$. In order to quantify this, we define the steady-state value $\mathcal{I}(\infty)$ by averaging over the last 50 time steps and study its behavior as a function of the system size; see Fig.~\ref{fig:figure_5}(b). One observes that for small (large) $W$, $\mathcal{I}(\infty)$ decreases (increases) with $L$, hinting at a vanishing (finite) value and thus an ergodic (localized) regime. The crossover happens around $W\sim 4$, which is in agreement with the results of Fig.~\ref{fig:figure_2}.

\subsection{Entanglement entropy}\label{subsec:entanglement}

Finally, we study the evolution of the entanglement entropy in the localized regime. We prepare the system in a random initial product state
\begin{equation}
|\psi_0\rangle = |n_{1},n_{2},...,n_{L}\rangle,
\end{equation}
where $n_{j}\in\{0,1,2\}$ is initialized randomly with the constraint $\sum_{j=1}^{L}n_{j}=N$. Using tensor networks, we compute the half-chain entanglement entropy
\begin{eqnarray}\label{eq:S_def}
S=-\text{tr}(\rho_{A}\ln \rho_A )= -\text{tr}(\rho_{B}\ln \rho_B ), 
\end{eqnarray}
where $\rho_{A,B}$ are the reduced density matrices of the left and right half of the chain. In an Anderson-localized phase $S$ becomes constant while in an MBL regime, $S$ grows logarithmically (i.e., without bounds) until finite-size effects set in \cite{Bardarson2012,Serbyn2013,Znidaric2008}. The entanglement dynamics can thus be used to discern Anderson from many-body localization.

\begin{figure}[t!]
\includegraphics[scale=0.29]{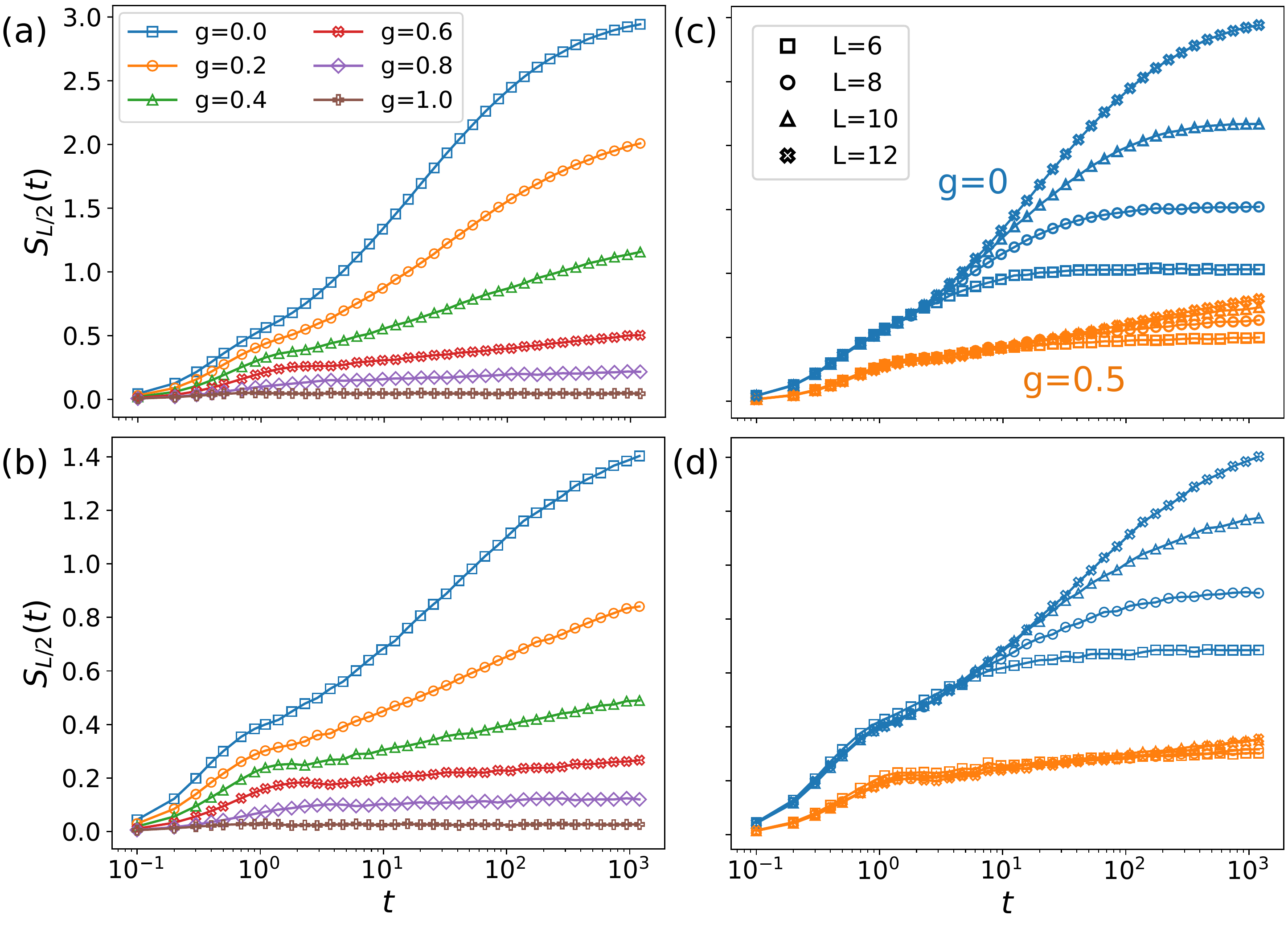}
\caption{Dynamics of the entanglement entropy in random product states at (a),(c) $W=5.0$, and (b),(d) $W=7.5$. (a) and (b) show data for fixed $L=12$ and various values of $g$; (c) and (d) illustrate the scaling with respect to the system size. In particular, we observe growth in the purely quadratic model ($g=0$), indicating many-body instead of Anderson localization.} 
\label{fig:figure_6}
\end{figure}

In Fig.~\ref{fig:figure_6}, we show the entanglement dynamics for various $g$ at $W=5$ [Figs.~\ref{fig:figure_6}(a) and~\ref{fig:figure_6}(c) ] and $W=7.5$ [Figs.~\ref{fig:figure_6}(b) and~\ref{fig:figure_6}(d) ]. For fixed $L=12$ (left column), $S$ grows logarithmically for all $g<1$, indicating that the system is many-body localized. This includes the point $g=0$ where the Hamiltonian is purely quadratic but not free due to the nontrivial statistics of the parafermions. At $g=1$, the entanglement does not grow; the system decomposes into segments of \emph{XX} chains, which, since they are free, become Anderson localized. The scaling with respect to the system size is shown in the right column of Fig.~\ref{fig:figure_6}, indicating that the logarithmic growth is only cut off by the finite system.

In sum, the entanglement dynamics in our model is similar to the one in the \emph{XXZ} chain if one identifies the point $g=1$ with the \emph{XX} limit.

\section{Conclusion}
\label{sec:conclusion}

We have studied the effects of disorder in a one-dimensional chain of Fock parafermions using both exact diagonalization and tensor network techniques. We calculated prototypical quantities such as level statistics or imbalance dynamics to demonstrate that disorder drives the system from an ergodic into a localized regime. Our results are representative for finite systems of ${\sim 14}$ sites, and we cannot rule out that the localized regime is, in fact, a prethermal precursor of a ergodic phase. The entanglement dynamics indicates that even a purely quadratic Fock parafermion model does not show Anderson but rather many-body localization due to the nontrivial statistics of the particles.

Our results indicate that a disordered parafermion model shows the same MBL phenomenology as the prototypical \emph{XXZ} chain; the point $g=1$ takes the role of the noninteracting \emph{XX} limit. The only difference is that in our case, the ergodic regime is generically described by a GUE instead of a GOE ensemble. The parafermion statistics seems to play the role of interactions in the \emph{XXZ} chain.  Whether or not disorder effectively removes all peculiarities of the parafermions is unclear and left for future work.

\emph{Note added.} We note that the pure tight-binding Fock parafermion chain ($g=0$) with disordered potential has been studied in Ref.~\cite{GrisevMBL} as well. As far as our studies overlap, all results are in qualitative agreement with our findings. One should note that in contrast to Ref. \cite{GrisevMBL} we do not impose the
condition $\sum_{j=1}^L \mu_j=0$, so a quantitative comparison of the
results is not possible.

\section*{Acknowledgements}
We would like to thank Vladimir Gritsev, Fabian Hassler, and Davide Rossini for useful discussions. Support by the Emmy Noether program of the Deutsche Forschungsgemeinschaft is acknowledged (Grant No. KA 3360/2-1). We acknowledge support by ``Niedersächsisches Vorab" through the Quantum- and Nano-Metrology (QUANOMET) initiative within the project P-1.  This work is part of the D-ITP consortium, a program of the Dutch Research Council (NWO) that is funded by the Dutch Ministry of Education, Culture and Science (OCW).

\bibliographystyle{apsrev4-2}
\bibliography{the_bibliography.bib}
 
\end{document}